\documentclass{article}

\usepackage{amsmath,amsfonts,amsmath,amsthm}
\usepackage{fullpage}
\usepackage{graphicx}
\usepackage{wasysym}
\usepackage{hyperref}
\usepackage{xcolor}

\newcommand{\tmix}{t_{\mathrm{mix}}}
\newcommand{\dmax}{d_{\max}}
\newcommand{\diam}{\mathrm{diam}}
\newcommand{\eee}{\mathrm{e}}
\newcommand{\Exp}{\mathbb{E}}
\newcommand{\Prob}{\mathbb{P}}
\newcommand{\dx}{\mathrm{d}x}
\newcommand{\dy}{\mathrm{d}y}
\newcommand{\ds}{\mathrm{d}s}
\newcommand{\dnew}{d_{\mathrm{new}}}
\newcommand{\poly}{\mathrm{poly}}

\newtheorem{theorem}{Theorem}
\newtheorem{remark}{Remark}

\def\tom#1{\marginpar{$\leftarrow$\fbox{*}}\footnote{$\Rightarrow$~{\sf #1 --Tom}}}
\newcommand{\cris}[1]{{\color{black} #1}}

\begin{document}

\title{Lower Bounds on the Critical Density in the Hard Disk Model via Optimized Metrics} 
\author{Thomas P. Hayes\thanks{Department of Computer Science, University of New Mexico. \texttt{hayes@cs.unm.edu} } 
\and Cristopher Moore\thanks{Santa Fe Institute. \texttt{moore@santafe.edu}}}
\maketitle

\begin{abstract}
We prove a new lower bound on the critical density $\rho_c$ of the hard disk model, i.e., 
the density below which it is possible to efficiently sample 
random configurations of $n$ non-overlapping disks in a unit torus.  
We use a classic Markov chain which moves one disk at a time, 
but with an improved path coupling analysis. 
Our main tool is an optimized metric on neighboring pairs of 
configurations, i.e., configurations that differ in the position of a single disk: 
we define a metric that depends on the difference in these positions, 
and which approaches zero continuously as they coincide.  
This improves the previous lower bound $\rho_c \ge 1/8$ to 
$\rho_c \ge 0.154$.
\end{abstract}

\section{Introduction}

The hard disk model is one of the simplest physical models of a non-ideal gas; yet it displays surprisingly complex behavior,
including multiple phase transitions.   A \emph{configuration} in this model consists of 
an arrangement of $n$ disks of radius $r$ on a unit
torus, so that no two of the disks overlap.  
At sufficiently high density, 
configurations tend to have a solid, or crystalline form.
At sufficiently low densities, there is a gaseous phase,
in which individual disks have plenty of room to move
around.  A more complex \emph{hexatic} phase has been observed in a narrow range of intermediate
densities~\cite{bernard2009event}.

Markov chain Monte Carlo algorithms have been used extensively to sample from this model's configuration space.  
Indeed, it was precisely this system that was studied in the classic paper of Metropolis et al.~\cite{metropolis}, who used Los Alamos' MANIAC computer to study a system with $n=224$ disks. 
The simplest such algorithms choose a disk and attempt to move it to a new position.  These algorithms work well at low densities, 
where the proposed position doesn't overlap with the other disks too often.  However, as with many other MCMC algorithms, our theoretical
guarantees lag well behind practice.  While it is believed that these algorithms mix in polynomial time throughout the gaseous 
phase, we can only prove this up to a density well below the experimental critical point.

The present work gives improved upper bounds on the mixing time for \emph{single-disk global-move dynamics}, which attempts to move 
a uniformly random disk to a uniformly random position anywhere in the torus.  As a corollary, we obtain a new rigorous lower bound on the 
critical density at which the gaseous phase ends.  

One of the challenging aspects of this particular Markov chain is that its state space is \emph{continuous}, whereas many of the existing techniques for analyzing mixing times of Markov chains are fundamentally discrete.  In particular, previous work~\cite{KMM-hard-core} used path coupling with the Hamming metric, where the distance between two configurations is $1$ whenever they differ in the position of a single disk.  Our approach is to define a metric that depends continuously on the difference between the two positions.  In particular, our metric goes to zero continuously as the two positions coincide, fitting our intuition that two configurations should be very close if a single disk has only been displaced a small amount.

\section{Preliminaries}
\subsection{Configurations of Hard Disks}

Let $\Omega$ denote the set of $n$-tuples of points in the
$d$-dimensional unit torus, $[0,1]^d$, that satisfy the constraint
that no two of the points are at distance less than $2r$ from one
another.  For now, let us assume $n$ and $r$ are such that $\Omega$
is non-empty, and moreover, of positive Lebesgue measure.
As a measurable subset of $[0,1]^d$, $\Omega$ inherits
the Lebesgue measure, and hence can be viewed as a probability space.
We will refer to this as the ``uniform distribution.''

Alternatively but equivalently, we can view each element of $\Omega$ as 
an arrangement, or configuration, of $n$ labelled, non-overlapping disks of radius $r$.
The combined volume of these disks is $\rho = n f(d,r)$ where $f(d,r)$
denotes the volume of a $d$-dimensional ball of radius $r$.  This can be 
given in closed form as
\[
f(d,r) = \frac{\pi^{d/2} r^d}{\Gamma(d/2 + 1)} \, .
\]
We will focus on the case $d=2$, where $f(d,r) = \pi r^2$.  

Since the volume of the torus $[0,1]^d$ is one, we call $\rho$ the
\emph{density} of each configuration in $\Omega$.  
A two-dimensional close packing of disks has density 
\[
\rho_{\rm packing} = \frac{\pi \sqrt{3}}{6} \approx 0.9069 \, .
\]
At this point the disks form a triangular lattice, and the state is a perfect crystal.  In contrast, at low enough densities the system is in a liquid or gaseous phase, where the disks have enough space between them to move freely.  Numerical results (e.g.~\cite{mak,piasecki2010prediction}) suggest a phase transition to a solid with long-range correlations at a critical density $\rho_c \approx 0.7$.  In addition, in between the solid and liquid phases there appears to be a \emph{hexatic} phase, with a rather subtle type of correlation.  Namely, if we draw lines between nearby pairs of disks, the orientations of these lines are correlated at large distances, even though the disks' positions are not~\cite{bernard2009event,bernard2011two,engel2013hard}.

\subsection{The Single-Particle Global-Move Dynamics}

Viewing the elements of $\Omega$ as arrangements of $n$ disks, there
is a natural Markov chain on $\Omega$ that moves at most one disk at
each timestep.  Namely, choose a disk at random, and a uniformly
random location in $[0,1]^d$.  If it would result in a valid
configuration, remove the chosen disk, and replace it by a new
disk centered at the chosen location.  If not, because the chosen
location is within distance $2r$ of one or more of the other disk
centers, then reject the move and leave the configuration unchanged.
This Markov chain is known as the \emph{single-particle global-move dynamics}.

It is easy to see that this Markov chain is reversible, and hence
the uniform distribution is stationary.  
The main question we are interested in is: for what densities $\rho$ does this Markov chain mix rapidly, in $\poly(n)$ steps?  
In physical terms, up to what density can it be proved that the system is in its gaseous phase?

It is worth mentioning that a number of other dynamics have been
proposed for $\Omega$, many of which are based on local moves
of one or more disks at a time.  The most efficient seems to be a non-reversible Markov chain called \emph{event chain dynamics}, 
due to Bernard, Krauth, and Wilson~\cite{bernard2009event}.  Unfortunately, we have even less rigorous understanding of these other dynamics than of single-disk dynamics.  

Kannan, Mahoney, and Montenegro~\cite[Theorem 2]{KMM-hard-core} proved the following.
\begin{theorem}  For $n$ hard-core particles in dimension $d$, the 
single-disk global-move dynamics has mixing time $\tau$ bounded by
\[
\tau = \begin{cases}
O(n \log n) & \mbox{if $2^{d+1} \rho < 1 - \gamma$ for any constant
  $\gamma$} \\
O(n^2 \log n) & \mbox{if $2^d \rho = 1$}.
\end{cases}
\]
\end{theorem}
\noindent
In particular, this shows that the critical density marking the end of the gaseous phase is bounded below by $\rho_c \ge 2^{-(d+1)}$, or $\rho_c \ge 1/8$ for $d=2$.

We review the proof of~\cite{KMM-hard-core}, which is based on path coupling~\cite{bubley-dyer}.  
Assume that $X_t, Y_t \in \Omega$ differ only in the position of a single disk, say disk 1.
Couple the dynamics so that, in the next timestep, both $X$ and $Y$ 
make the same choice of disk $j$ to move.  If $j = 1$, which occurs with probability $1/n$, 
we choose the same proposed position in both chains.
The chains will coalesce if the proposed position is valid; otherwise,
the Hamming distance remains 1.  Since a fraction $\rho$ of the total volume is 
within distance $r$ of a center, a fraction at most $2^d \rho$ is within
distance $2r$ of a center.  Thus coalescence occurs with
probability at least $(1 - 2^d \rho)/n$.

If $j \ne 1$, we define the coupling so that the proposed position is identical if
it is valid in both chains.  Otherwise, we choose the proposed position uniformly for $X$, 
and obtain the proposed position for $Y$ by reflecting around the 
line bisecting the two positions of disk 1.  
This guarantees that for at least half of these scenarios, the attempted update will fail in
both chains.  
\cris{(This is reminiscent of Jerrum's coupling for graph coloring, 
which matches forbidden colors with each other to increase the probability that a move will be rejected in both chains~\cite{jerrum}.)}
For the other half, one or both updates may succeed, increasing
the Hamming distance to 2.  Since a $\rho/n$ fraction of the total volume
is covered by disk 1, the probability of the Hamming distance
increasing is at most $2^d \rho/n$.

Comparing these probabilities, we see that the expected change in
Hamming distance $d(X_t,Y_t)$ is negative whenever $1 - 2^d \rho > 2^d \rho$, 
or equivalently, when $\rho < 2^{-(d+1)}$.  When $\rho = 2^{-(d+1)}$, 
the proof is completed by lower bounding the expected squared change 
in $d(X_t,Y_t)$.


\section{Related Work}

Our work is different than most previous work on Markov chains in two respects.  First, the state space is continuous.  Second, we improve a path coupling argument not by optimizing the coupling or expanding the set of moves, but by optimizing the metric we use to define the distance between neighboring configurations. 

Markov chains with continuous state spaces have received relatively little attention in computer science.  An important recent exception is Randall and Winkler~\cite{randall2005mixing-int,randall2005mixing-circ}, who analyzed the problem of arranging $n$ dots in an interval, or on a
circle.  Indeed, their work can be thought of as the one-dimensional version of the hard disk model.

Vigoda~\cite{vigoda2001note} was one of the first to use an adaptively
weighted Hamming metric to improve a path coupling argument.
Earlier work by Dyer and Greenhill~\cite{dyer2000markov} and Luby and
Vigoda~\cite{luby1999fast} had established that the Glauber dynamics for the hard-core model
on graphs of maximum degree $\Delta$ mixes in polynomial time 
when the fugacity $\lambda$ is bounded below $2/(\Delta - 2)$.  
Both of these works proceeded by proving $O(n \log n)$ mixing for a related Markov chain with an expanded set of moves, 
which they related to Glauber dynamics with a comparison argument: 
namely, they added  a ``slide'' transition that can move a particle between two
adjacent sites.  By suitably balancing the probabilities for
single-site updates against those for the slide moves, they showed that a greedy
coupling is contractive for the Hamming metric.

Rather than adding a slide transition or altering the coupling, Vigoda~\cite{vigoda2001note} obtained the same bound by defining a metric in which a single disagreement counts as distance $1 - c|B_v|$ where $B_v$ denotes the set of ``blocked'' neighbors of the disagreeing vertex $v$, and $c$ is a suitably chosen parameter.  (A vertex is blocked if any of its neighbors, excluding $v$, is occupied.)  Each such blocked neighbor corresponds to a move which will be rejected rather than driving the two configurations apart.  Thus Vigoda's metric ``rewards'' pairs of configurations, considering them closer than the Hamming metric would suggest, whenever the probability that the simple coupling would increase the disagreement them is relatively low.  Our approach is similar in spirit: we define the distance between two neighboring configurations as small whenever the disagreeing disk is only displaced slightly from one configuration to the other.

\section{Main Result}

We prove the following.
\begin{theorem}
\label{thm:main}
For $n$ hard-core particles in $d=2$ dimensions, the single-disk global-move dynamics has mixing time $\tau = O(n \log n)$ 
whenever 
\[
\rho \le 0.154483...
\]
Thus the critical density below which the system is in its gaseous phase is bounded by $\rho_c \ge 0.154483...$
\end{theorem}
\noindent
\cris{Our results represents a modest improvement over the previous
  lower bound on the critical density, $\rho_c \ge 1/8$, and remain far below the experimental value $\rho_c \approx 0.7$.  However, we believe that our bound can be improved further by combining our continuous metric with other geometrical ideas.}

Like~\cite{KMM-hard-core}, our proof works by path coupling.  Indeed, we use precisely the same coupling they do; 
our improvement is entirely due to the choice of metric.  
For two configurations $X, Y$ with Hamming distance $1$, i.e., that disagree only in the location of one disk, 
if the two differing locations are at a Euclidean distance $\ell$,
we define the distance between the two configurations as $D(X,Y) = d(\ell)$,
for a suitable subadditive function $d$ with 
\cris{$d(\ell) = \dmax = 1$ for $\ell$ sufficiently large.}  
We then optimize the function $d(\ell)$ to obtain the largest possible $\rho$ such that 
the coupling is contractive, i.e., so that $\Exp[\Delta d(X,Y)]$ is bounded below zero for all pairs $X, Y$ with Hamming distance $1$.

As usual in path coupling, we extend this metric to pairs $X, Y$ with Hamming distance $h > 1$ by considering paths through $\Omega$ of length $h$: that is, $Z_0 = X, Z_1, Z_2, \dots, Z_{h-1}, Z_h = Y$, where each pair $(Z_i, Z_{i+1})$
differ in the position of a single disk.  Then we define
\[
D(X,Y) = \min \sum_{i=0}^{h-1} D(Z_i, Z_{i+1}) 
\]
where the minimum is taken over all the possible paths $Z_1, \dots, Z_{h-1}$.  If $\Exp[\Delta d(X,Y)]$ is bounded below zero for all $X, Y$ with Hamming distance $1$, it follows that $\Exp[\Delta d(X,Y)] < 0$ for all $(X, Y) \in \Omega \times \Omega$.  Since the maximum Hamming distance is $n$, this is enough to establish mixing in $O(n \log n)$ time.

There are two differences between our metric and that used in~\cite{KMM-hard-core}.  The first is that our function
$d(\ell)$ depends continuously on $\ell$, and $d(\ell) \to 0$ as $\ell \to 0$.  In contrast, they used Hamming distance, where $d(\ell) = 1$ for all $\ell > 0$.  

Secondly, their notion of Hamming distance is slightly different.  They use a fixed labelling of the $n$ disks, so that Hamming distance means the number of disks $i$ whose positions differ in the two configurations.  However, when we speak of the Hamming distance between two configurations, we always think of unlabelled or adaptively labelled disks, so that as few disks as possible are in differing positions under the two configurations.  In general, these adaptively chosen labellings will change from timestep to timestep as the configurations evolve.  
In particular, if two configurations differ only in that two disks
have been switched, we consider them identical rather than having
Hamming distance $2$.  This further reduces the shortest-path metric
between two configurations that differ on two disks; if disk $i$ in
$X$ is close to disk $j$ in $Y$ and vice versa, our metric considers
$X$ and $Y$ to be close, even if $i$ and $j$ are distant from each
other in both configurations.
As we will see, it is this switching of two labels that lets us take
advantage of our continuous metric to prove contractivity at a higher density.


\subsection{Path Coupling in Continuous State Spaces}

The following version of Bubley and Dyer's path-coupling theorem 
is taken from the textbook of Levin, Peres and
Wilmer~\cite[Theorem 14.6, Corollary 14.7]{LPW}.

\begin{theorem}
Suppose the state space $\Omega$ of a Markov chain is the vertex set
of a graph with length function $d \ge 1$ defined on edges.
Let $\rho$ be the corresponding shortest-path metric.
Suppose that for each edge $\{x,y\}$, there exists a coupling
$(X_1, Y_1)$ of the distributions $P(x, \cdot)$, $P(y, \cdot)$
such that 
\[
\Exp_{x,y} [ \rho(X_1, Y_1) ] \le \rho(x,y) \,\eee^{-\alpha} 
= d(x,y) \,\eee^{-\alpha}.
\]
Then 
\[
\tmix(\epsilon) \le \left\lceil \frac{ - \log(\epsilon) +
    \log(\diam(\Omega))}{\alpha} \right \rceil .
\]
\end{theorem}

In our continuous setting, the length function
$d$ that is not bounded away from zero, so we cannot apply this theorem directly.  
However, the idea behind its proof immediately yields the following variant.

\begin{theorem} \label{thm:path-coupling-variant}
Suppose the state space $\Omega$ of a Markov chain is the vertex set
of a graph with length function $d$ defined on edges.
Let $\rho$ be the corresponding shortest-path metric.
Suppose that for each edge $\{x,y\}$, there exists a coupling
$(X_1, Y_1)$ of the distributions $P(x, \cdot)$, $P(y, \cdot)$
such that 
\[
\Exp_{x,y} ( \rho(X_1, Y_1) ) \le \rho(x,y) \,\eee^{-\alpha} 
= d(x,y) \,\eee^{-\alpha}.
\]
Then, for  all $X_0, Y_0$, all $\epsilon > 0$ and all $t \ge 1$,
\[
\Prob[ \rho(X_t, Y_t) > \epsilon ] \le \diam(\Omega) \,\eee^{-\alpha t}.
\]
\end{theorem}

Once we have a pair of configurations whose distance is sufficiently
small, say $\epsilon/(n \log n)$,
we can argue that over the course of the next $O(n \log n)$ updates, every disk
that is successfully moved in one chain is moved to the same position
in the other chain.  By coupon collecting, this is enough steps to
almost surely move each disk, so the two chains have
coalesced with probability $1 - O(\epsilon)$.
Combining this with Theorem~\ref{thm:path-coupling-variant}, 
setting $\epsilon = O(1/(n \log n))$, $\diam(\Omega) = O(n)$, $\alpha = O(1/n)$, we obtain a mixing time of $O(n \log n)$.

The above argument is not new; it was used by Randall and
Winkler~\cite[Lemma 3]{randall2005mixing-int} to analyze a local-move 
Markov chain for a one-dimensional version of the hard-core model.

\subsection{Analysis of the Path Coupling}

\begin{figure}
\begin{center}
\includegraphics[width=\textwidth]{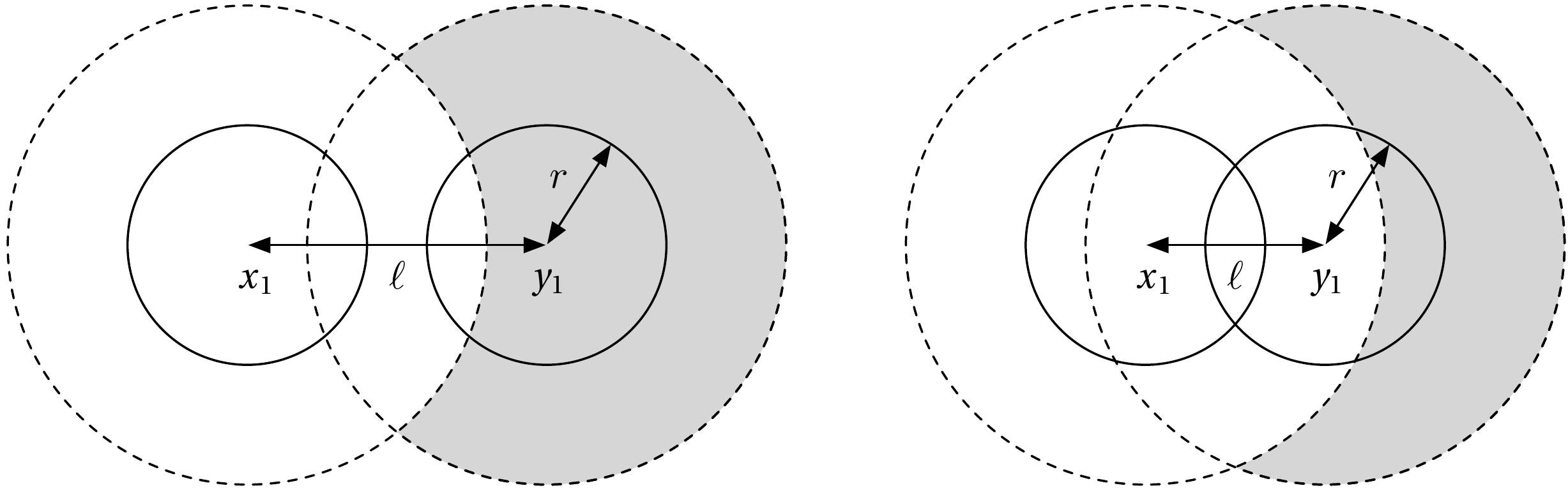}
\end{center}
\caption{A disk centered at $x_j$ is surrounded by a ``danger zone'' $Z(x_j)$ of radius $2r$.  If two configurations differ at disk $1$, a bad move only occurs if we try to move a disk into the ``danger crescent'' (shaded) $Z(y_1) \setminus Z(x_1)$, which we couple to its mirror image.  On the left, $\ell > 2r$ and the crescent includes the center $y_1$ of the disk on the right.  On the right, $\ell < 2r$ and the crescent excludes $y_1$.}
\label{fig:danger-zones}
\end{figure}

\begin{figure}
\begin{center}
\includegraphics[width=0.5\textwidth]{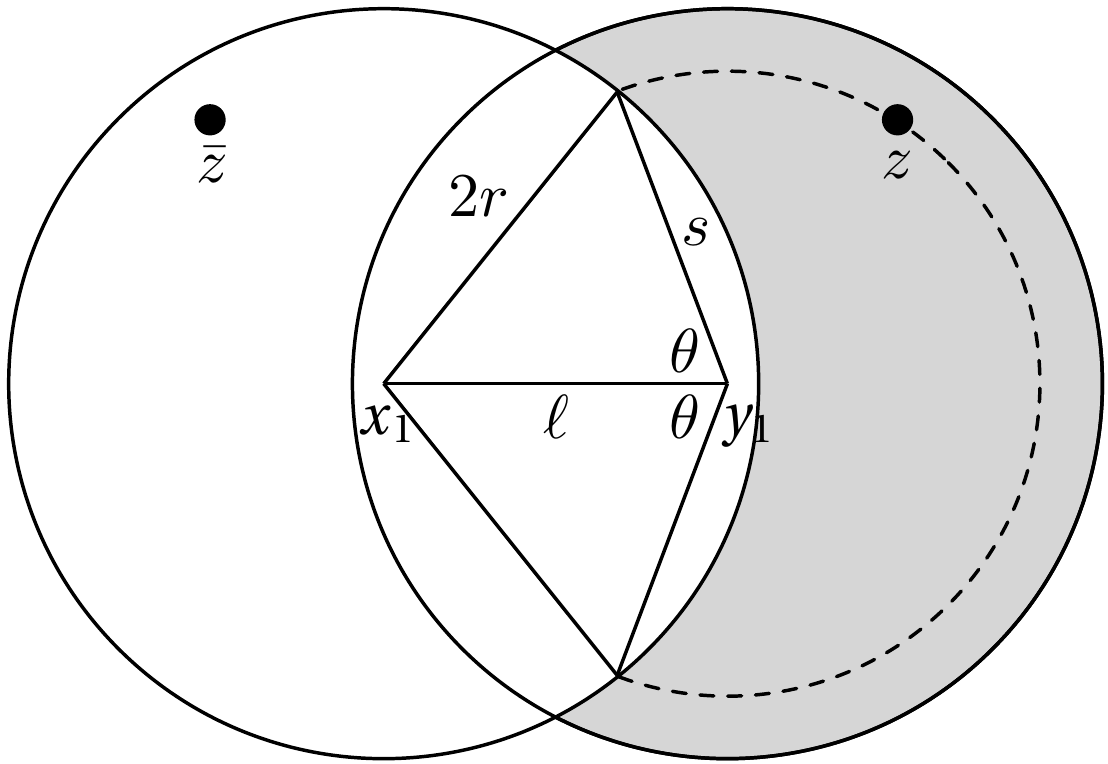}
\end{center}
\caption{We attempt to move a disk to $z$ in the $X$ chain, and to its mirror image $\bar{z}$ in the $Y$ chain.  If $\|z-y_1\| = s$, then the danger crescent (shaded) ranges over a circular arc (dashed) centered at $y_1$ and subtending an angle $2(\pi-\theta)$ where $\theta = \theta(s,\ell)$ is given by~\eqref{eq:theta}.}
\label{fig:theta}
\end{figure}

Let $X, Y$ be two configurations differing in the position of
one disk, which we call disk $1$.  Denote the disks' centers as $X = (x_1, x_2, \dots, x_n)$
and $Y = (y_1, y_2, \dots, y_n)$, where $x_1 \ne y_1$ and $x_j = y_j$
for all $j \ge 2$.  We write $\ell = \| x_1 - y_1 \|$, so the distance
between $X$ and $Y$ in our metric is $d(\ell)$.

If disk $1$ is chosen for the move, then the distance will decrease to
zero if the proposed position is legal.  
Each disk $x_j$ is surrounded by a ``danger zone'' $Z(x_j)$ of 
radius $2r$, which has area $4\pi r^2$.  
As in~\cite{KMM-hard-core}, 
we pessimistically assume that these danger zones are disjoint; 
equivalently, we use the union bound for the probability that the proposed position 
falls into any of them.  Thus $X$ and $Y$ coalesce with probability at least $1 - 4\pi r^2 n = 1 - 4 \rho$.  

If any other disk $j \ge 2$ is chosen for the move, then the update will
ordinarily succeed or fail in both chains.  The one case where it
may not is when the proposed position $z$ is in the danger zone of $y_1$ but outside the danger zone of $x_1$.  
We will call this set $Z(y_1) \setminus Z(x_1)$ the 
\emph{danger crescent}, or simply the ``crescent,''
since, at least when $d=2$, its shape is known as a crescent or lune.

When the $X$ chain proposes a position $z$ in the danger crescent,
the $Y$ chain proposes the mirror image position in
$Z(x_1) \setminus Z(y_1)$, which we denote $\bar{z}$.    
If the update succeeds in either or both chains, our metric will generally change, 
either increasing or decreasing, as we will investigate more closely below.

The above cases are the only ways that $d(X,Y)$ can change.
In particular, note that when $X$ proposes a move into the mirror
image of the crescent, the proposed move is always rejected in both
chains, being blocked by disk $1$ in both cases.

When one or both updates succeed, we need to decide whether
the two disks formerly labelled ``disk $j$'' should remain paired
together, or whether one of them should swap places with disk $1$
in our pairing.  This is done greedily, minimizing
the new distance between the configurations.


As shown in Fig.~\ref{fig:theta}, let $s = \|z - y_1\|$.  
We will distinguish two cases.  
When $s \ge \ell$ and the move succeeds in at least one of the two chains, we call it a ``far move.''  In this case we keep the original indexing of the disks, so that on the next step we have $x'_1 = x_1$ and $x'_j = z$ if the move succeeds in the $X$ chain, and similarly in the $Y$ chain.  Since the $x_j$ is distant from $x_1$, $y_1$, and $z$ with probability $1-O(1/n)$, we use the trivial upper bound $d(x'_j,y'_j) \le \dmax = 1$ on the change in distance.

When $s < \ell$ and the move succeeds in at least one of the two chains, we call it a ``near move.''  Suppose without loss of generality that the move succeeds in the $X$ chain, i.e., $z$ is not blocked in the $X$ chain.  In this case, we swap the labels $1$ and $j$ in the $X$ chain, setting $x'_1 = z$ and $x'_j = x_1$.  Since the actual distance is the minimum over all relabelings, this choice, like any other, will imply an upper bound on the expected distance.  There are then two cases: either the mirror image $\bar{z}$ is blocked in the $Y$ chain or not.  If it is not, so that the move succeeds in both chains, we set $y'_j = \bar{z}$; in either case, we set $y'_1 = y_1$.   Thus the change in our metric is at most 
\begin{align*}
& d(\|x'_j-y'_j\|) + d(\|x'_1 - y'_1\|) - d(\| x_1-y_1 \|) \\
\le\;& \dmax + d(\|z-y_1\|) - d(\| x_1-y_1 \|) \\
=\;& \dmax + d( s ) - d(\ell) \, , 
\end{align*}
where again $\dmax = 1$.  

Combining the above formulas, our upper bound on the expected change in the metric due to new disagreements becomes $\dmax$ times the area of the danger crescent, minus the integral of the potential ``savings'' $d(\ell) - d(s)$ over the subregion $s \le \ell$ corresponding to near moves.

Now, using a formula for the area of intersection of two circles,
we find the area of the danger crescent equals
\begin{equation}
\label{eq:amoon}
A_{\!\rightmoon}(\ell) = r^2 \left( 8 \sin^{-1} \left(
    \frac{\ell}{4r}\right) + \frac{\ell}{r} \sqrt{ 4  -
    \frac{\ell^2}{4r^2} } \right).
\end{equation}
Integrating the savings $d(\ell) - d(s)$ is more difficult, because
we don't know the right choice of $d$.
However, we can still set up the integral.  For $0 \le s \le \ell$,
let $\theta(s, \ell)$ denote the angle shown in Fig.~\ref{fig:theta}.  
When the triple $(s, \ell, 2r)$ satisfies the triangle
inequality, $\theta$ is given by the law of cosines:
\begin{equation}
\label{eq:theta}
\theta(s,\ell) = \cos^{-1} \left( \frac{s^2 + \ell^2 - 4r^2}{2 \ell s} \right)
\end{equation}
\cris{Otherwise, let $\theta = 0$ when $s < \ell - 2r$, and $\theta = \pi$ when $s < 2r - \ell$.}

Note that when $\ell \le r$, we have $\theta(s, \ell) = \pi$ for all $s
\le \ell$.  This means there is never any savings in this case, so 
the correct setting of $d(\ell)$ for $0 \le \ell \le r$ satisfies
\begin{equation}
\label{eq:tom}
d(\ell) = \frac{n}{1 - 4\rho} A_{\!\rightmoon}(\ell)
= \frac{\rho}{\pi(1 - 4 \rho)}  \left( 8 \sin^{-1} \left(
    \frac{\ell}{4r}\right) + \frac{\ell}{r} \sqrt{ 4  -
    \frac{\ell^2}{4r^2} } \right).
\end{equation}
So, in the interval $0 \le \ell \le r$ we can determine the optimal metric analytically.

Now, integrating in polar coordinates, the expected savings equals
\[
\int_{s = 0}^{\ell} 2(\pi-\theta(s,\ell)) \,(d(\ell) - d(s)) \,s \; \ds \,.
\]
Putting this all together, our coupling will be $\epsilon$-contractive relative
to our chosen metric if, for all $0 < \ell \le 4r$, we have
\begin{equation} 
\label{eq:linear-system-integrals}
\left( \frac{1 - 4\rho}{n} - \epsilon \right) d(\ell) \ge
A_{\!\rightmoon}(\ell) - \int_{0}^{\ell} 2(\pi-\theta(s,\ell)) \,(d(\ell) - d(s)) \,s \; \ds \,.
\end{equation}
We set $d(\ell) = 1$ for all $\ell \ge 4r$.

\begin{figure}
\begin{center}
\includegraphics[width=0.5\textwidth]{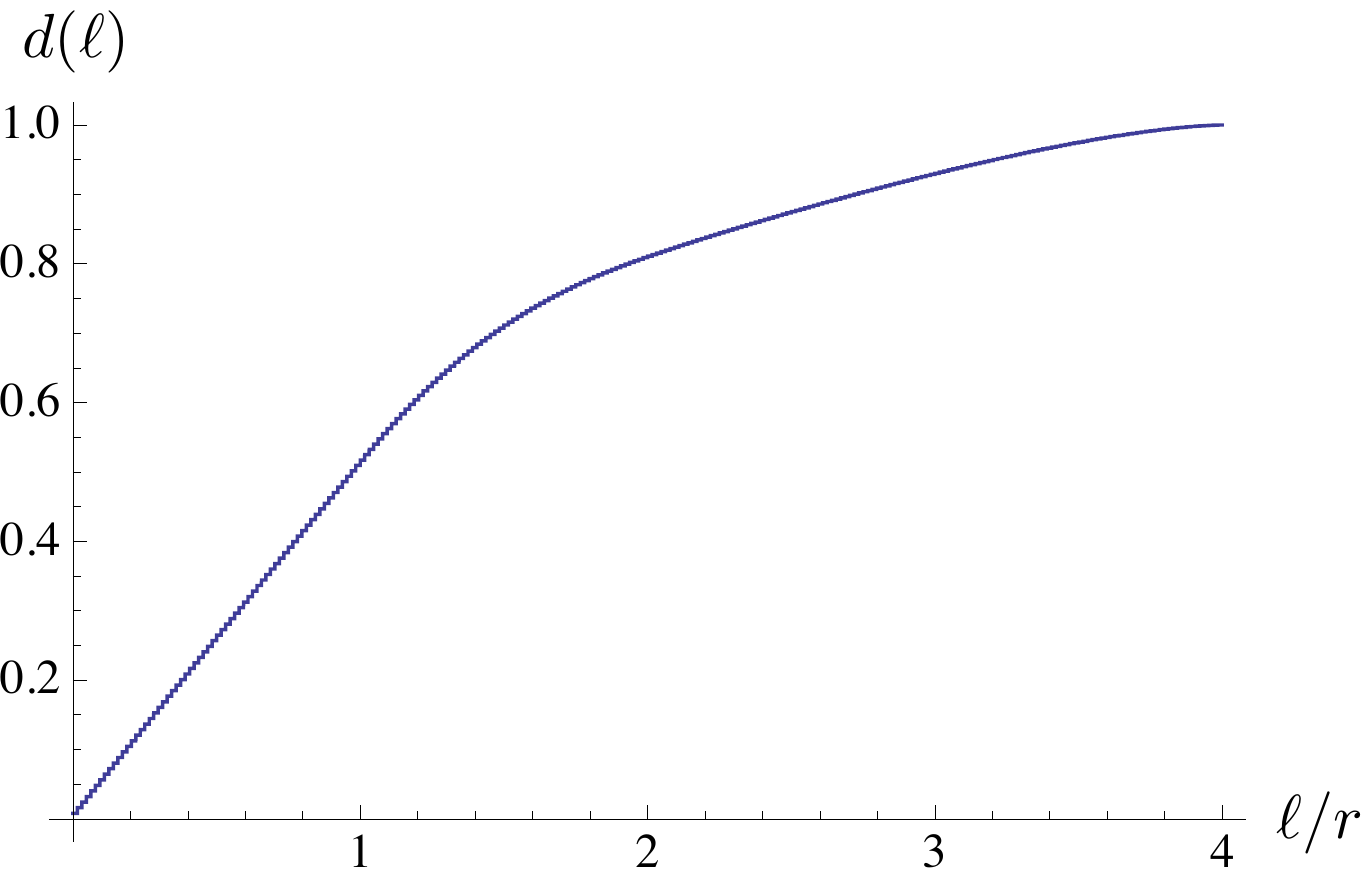}
\end{center}
\caption{The optimal metric resulting from dividing $[0,4]$ into $L=256$ equal subintervals, setting $d(\ell)$ equal to a constant in each subinterval, and solving the resulting LP.}
\label{fig:metric}
\end{figure}

\cris{Now, for any fixed choice of $\rho$ and $\epsilon$, 
we can consider various families of metrics.  In particular, 
if we divide the interval $[0,4r]$ into $L$ subintervals of width $4r/L$ each, 
and assume that $d(\ell)$ is constant within each subinterval, the 
integral in~\eqref{eq:linear-system-integrals} becomes a finite Riemann sum.    
For each $\rho$ this gives a linear program with $m$ variables.  We then 
perform a binary search, finding the largest $\rho$ for which this LP is feasible.  
Setting $\epsilon=10^{-6}/n$ and using a numerical software package to determine LP feasibility, 
we obtain the results shown in Table~\ref{tab}.  It seems unlikely that increasing $L$ will improve the first three digits of $\rho_c$.  

The metric $d(\ell)$ we obtain is shown in Fig.~\ref{fig:metric}.  It appears to have a piecewise analytic form, but we have made only desultory attempts to determine it except for~\eqref{eq:tom} in the range $0 \le \ell \le r$ (with which there is excellent agreement).  Interestingly, the constraints in~\ref{eq:linear-system-integrals} seem to be tight for $0 \le \ell \le 2r$, and slack for $2r < \ell \le 4r$.
}

\begin{table}
\[
\label{tab}
\begin{array}{|c|cccccc|}
\hline
L & 8 & 16 & 32 & 64 & 128 & 256 \\
\rho_c & 
0.150024 & 
0.152182 & 
0.153373 & 
0.153999 &
0.154320 &
0.154483 \\ \hline
\end{array}
\]
\caption{Lower bounds on $\rho_c$ obtained by dividing the interval $[0,4r]$ into $L$ subintervals.}
\end{table}


\section{Further Improvement}

\cris{Our new lower bound on the critical density is still very far from the
experimental value $\rho_c \approx 0.7$.  Our analysis of the coupling of single-disk global-move 
dynamics can undoubtedly be improved significantly, even if we stick with the current coupling.  
It seems especially important to reduce the probability that a good move is rejected, 
i.e., reduce the term $4\rho$ on the left side of~\eqref{eq:linear-system-integrals}.  
Equivalently, we need to use the fact that, at least after a warm start, many of the danger zones overlap with high probability, 
so that the expected area of their union is less than $4\rho$.  We believe that this and other ideas will let us push the lower bound 
on $\rho_c$ above $1/4$.
}

Another direction for improvement would be to
increase the probability of savings from modifying
the pairing.  In particular, the worst case for our
current analysis has all new disagreements come
from moves that are blocked in exactly one of the 
two chains.   Perhaps
by accounting for the possibility that some of the 
blocking disks can be moved away, this
worst case could be avoided.  

\section*{Acknowledgments}  
We benefited from the Workshop on Disorder, Algorithms, and Complexity at the Aspen Center for Physics, and from conversations with Werner Krauth.  This work was supported by NSF grant CCF-1219117.

\bibliography{hard-disks}

\begin{thebibliography}{10}

\bibitem{bernard2011two}
Etienne~P Bernard and Werner Krauth.
\newblock Two-step melting in two dimensions: first-order liquid-hexatic
  transition.
\newblock {\em Physical Review Letters}, 107(15):155704, 2011.

\bibitem{bernard2009event}
Etienne~P Bernard, Werner Krauth, and David~B Wilson.
\newblock Event-chain monte carlo algorithms for hard-sphere systems.
\newblock {\em Physical Review E}, 80(5):056704, 2009.

\bibitem{bubley-dyer}
Russ Bubley and Martin~E. Dyer.
\newblock Path coupling: A technique for proving rapid mixing in {M}arkov
  chains.
\newblock In {\em Proc. FOCS}, pages 223--231, 1997.

\bibitem{dyer2000markov}
Martin Dyer and Catherine Greenhill.
\newblock On {M}arkov chains for independent sets.
\newblock {\em Journal of Algorithms}, 35(1):17--49, 2000.

\bibitem{engel2013hard}
Michael Engel, Joshua~A Anderson, Sharon~C Glotzer, Masaharu Isobe, Etienne~P
  Bernard, and Werner Krauth.
\newblock Hard-disk equation of state: First-order liquid-hexatic transition in
  two dimensions with three simulation methods.
\newblock {\em Physical Review E}, 87(4):042134, 2013.

\bibitem{jerrum}
Mark Jerrum.
\newblock A very simple algorithm for estimating the number of $k$-colourings
  of a low-degree graph.
\newblock {\em Random Structures and Algorithms}, 7:157--165, 1995.

\bibitem{KMM-hard-core}
Ravi Kannan, Michael~W. Mahoney, and Ravi Montenegro.
\newblock Rapid mixing of several {M}arkov chains for a hard-core model.
\newblock In {\em Proc. 14th Intl. Symp. on Algorithms and Computation
  (ISAAC)}, pages 663--675, 2003.

\bibitem{LPW}
David~A. Levin, Yuval Peres, and Elizabeth~L. Wilmer.
\newblock {\em {M}arkov Chains and Mixing Times}.
\newblock AMS, 2008.

\bibitem{luby1999fast}
Michael Luby and Eric Vigoda.
\newblock Fast convergence of the {G}lauber dynamics for sampling independent
  sets: Part {I}.
\newblock {\em Random Structures and Algorithms}, 15(3-4):229--241, 1999.

\bibitem{mak}
C.~H. Mak.
\newblock {Large-scale simulations of the two-dimensional melting of hard
  disks}.
\newblock {\em Phys. Rev. E}, 73(6):065104, 2006.

\bibitem{metropolis}
N.~Metropolis, A.~Rosenbluth, M.~Rosenbluth, A.~Teller, and E.~Teller.
\newblock Equation of state calculations by fast computing machines.
\newblock {\em J. Chem. Phys.}, 21, 1953.

\bibitem{piasecki2010prediction}
Jaros{\l}aw Piasecki, Piotr Szymczak, and John~J Kozak.
\newblock Prediction of a structural transition in the hard disk fluid.
\newblock {\em The Journal of chemical physics}, 133:164507, 2010.

\bibitem{randall2005mixing-circ}
Dana Randall and Peter Winkler.
\newblock Mixing points on a circle.
\newblock In {\em Approximation, Randomization and Combinatorial Optimization.
  Algorithms and Techniques}, pages 426--435. Springer, 2005.

\bibitem{randall2005mixing-int}
Dana Randall and Peter Winkler.
\newblock Mixing points on an interval.
\newblock In {\em Proc. Second Workshop on Analytic Algorthmics and
  Combinatorics (ANALCO)}, pages 218--221, 2005.

\bibitem{vigoda2001note}
Eric Vigoda.
\newblock A note on the {G}lauber dynamics for sampling independent sets.
\newblock {\em Electronic Journal of Combinatorics}, 8(1):1--8, 2001.

\end{thebibliography}

\end{document}